\documentclass[twocolumn,pra,showpacs]{revtex4}
\usepackage{graphicx}
\usepackage{dcolumn}
\usepackage{bm}
\preprint{APS/123-QED}
\begin{document}
\title{Electromagnetically induced left-handedness in a dense gas of three level atoms}
\author{M. \"O. Oktel }
\affiliation{Bilkent University, Department of Physics, 06800 Bilkent, Ankara, Turkey}
\email{oktel@fen.bilkent.edu.tr}
\author{\"O. E. M\"ustecapl{\i}o\u{g}lu}
\affiliation{Ko\c{c} University, Department of Physics,
Rumelifeneri Yolu, 34450 Sar{\i}yer, Istanbul, Turkey}
\email{omustecap@ku.edu.tr}
\date{\today}
\begin{abstract}
We discuss how a three level system can be used to change the frequency dependent
magnetic permeability of an atomic gas to be significantly different from one. We derive
the conditions for such a scheme to be successful and briefly discuss the resulting
macroscopic electrodynamics. We find that it may be possible to obtain left handed electrodynamics
for an atomic gas using three atomic levels.
\end{abstract}
\pacs{ 42.50.Gy,03.75.Nt, 42.25.Bs         }
\maketitle
\section{Introduction}
\label{sec:introduction}
Changing the propagation properties of light by designing a novel material  is of
interest both from a  basic science point of view, and for technological applications.
Recent advances such as slowing down\cite{slow} or stopping light\cite{stop}, or
left handed metamaterials \cite{physicstoday}
promise advances in fields ranging from optics \cite{focusing}  to quantum computation\cite{quantumcomputation}.
 It is desirable to find new materials
in which electromagnetic waves exhibit novel behavior, and there is a flurry of activity both theoretically and
experimentally in this direction.

The macroscopic electromagnetic (EM) properties of a medium are characterized by the frequency
dependent dielectric constant\cite{Jackson}
\begin{equation}
\vec{D} = \varepsilon(\omega) \vec{E}
\end{equation}
and the magnetic permeability
\begin{equation}
\vec{B} = \mu(\omega) \vec{H}.
\end{equation}

Speed of an EM wave of frequency $\omega$ in this medium is given by
\begin{equation}
\label{speedoflight}
v=\frac{c}{\sqrt{\mu(\omega) \varepsilon(\omega)}}
\end{equation}
where c is the speed of light in vacuum. The index of refraction is then
$n = \sqrt{\mu(\omega) \varepsilon(\omega)}.$

The dielectric constant of the medium shows large variations near a resonance, {\it i.e.} when the frequency of the
external field is near an internal state transition. This makes it easy to change the refractive index of a medium
by properly tuning the frequency of the EM wave to just above or below a transition.  Recently this fact was employed
combined with quantum coherent effects to get very high refractive indices in atomic gases \cite{3level,slow}.

Although $\varepsilon(\omega)$ can change appreciably for a gas of atoms at optical frequencies, the magnetic
permeability $\mu(\omega)$ is always very close to its free space value.
One can give arguments in classical electrodynamics to explain
this \cite{Landau}, or understand it in terms of atomic transitions as follows.
Magnetic field component of  an EM wave couples to the atom much weaker than the electric field component. The magnetic coupling to an atom is proportional to
the Bohr magneton $\mu_B=\frac{e \hbar}{2 m_e c} = \alpha e a_0$, while the electric coupling is $e a_0$.
The fine structure constant $\alpha \simeq 1/137$ also shows itself in the induced magnetic dipole moment.
Overall the effect of an EM
wave on magnetic permeability is $\alpha^2$ weaker than its effect on the electric susceptibility. Another important
fact is that magnetic dipole transitions are allowed only between states which have the same radial wavefunction, and
generally two such states are not separated by optical frequencies in energy.

Now that it is hard to get $\mu(\omega)$ to be different than one, we need to question why it is important to have
another value for it. After all it seems from Eq.(\ref{speedoflight}) that all the optical properties of the
medium depend on the product $ \mu(\omega) \varepsilon(\omega). $ The answer to this question is that the refractive
index alone does not completely represent the medium \cite{Veselago}. One can imagine two media, one with $\varepsilon_1(\omega) > 0 $, $ \mu_1(\omega) > 0 $ and $ \epsilon_2(\omega) = - \epsilon_1(\omega) $,
$ \mu_2(\omega) = - \mu_1(\omega) $. They would have the same refraction index, however, quite different
optical properties. Materials with both $\varepsilon < 0$ and $\mu < 0$, are called left handed materials,
named for parity of the coordinate frame formed by $\{\vec{E},\vec{H},\vec{k}\}$. Optical properties of left handed
and right handed materials differ mainly because, the Poynting vector points opposite to $\vec{k}$ in left handed
materials. Most remarkable change happens at the interface between a left handed material and a right
handed material, where the usual Snell's refraction law gets a sign change. In addition to the inverse Snell's law,
the reverse Cerenkov radiation and the reverse Doppler shift would also be possible in such materials\cite{Veselago}.

 Left handed artificial materials in the microwave region have recently been built\cite{leftexperiment}
by assembling
a composite lattice of metallic split ring resonators and metallic wires\cite{lefttheory},
with periodicity much smaller than the wavelength of the
electromagnetic field, or using anomalous propagation properties
of light in a photonic crystal, with periodicity is in the order of the wavelength of the electromagnetic
radiation\cite{ertugrul}. All such systems (called metamaterials), require delicate manufacturing of
spatially periodic structures.
In the microwave region, improvements of focusing, filtering and steering properties of microwaves would be
useful for many practical applications. Similar improvements would also be valuable for
 applications operating at optical frequencies. In this paper we examine the case of an atomic gas without
any spatial periodicity that could exhibit behavior similar to metamaterials at optical frequencies.

 We have remarked that the magnetic dipole response to an oscillating magnetic field is smaller by a
factor of $\alpha^2$ compared to the electric dipole response to an oscillating magnetic field.
In an EM wave $\vec{E}$ and $\vec{B}$ fields are always perpendicular to each other and are always in phase.
If one can get the atom to respond to an electric field $\vec{E}$ with a magnetic moment $\vec{\mu}$
perpendicular to it, and in phase with it, one can effectively think that the magnetic dipole moment
is induced by the magnetic field of the EM wave. Thus, it is possible get a magnetic response which is only $\alpha$
times smaller than the electric response. Such response enables one to achieve a regime where the propagation
properties of light are significantly different.

The aim of this paper is to explore the feasibility of this idea to modify the magnetic permeability  of an
atomic gas electromagnetically. To this end, we introduce a model system in the next section and find the necessary conditions for the
applicability of our scheme. In section III we present the results of our calculations for two different parameter
regimes, a dilute gas and a dense gas. We then go on to discuss the consequences of our results for experiments. Finally,
we give a summary of our results and conclusions in section IV.

\section{Model system}
\label{sec:system}

In this section, we construct a model system for which the magnetic permeability can be optically
modified. We also describe the scheme for modification in detail, and discuss its limitations.

One can readily conclude  by parity arguments that it is not possible to get a magnetic response to an
electric field if only two states are involved. An electric field causes transitions to states which are
of opposite parity to ground state, and such states do not have a magnetic dipole matrix element with
the ground state. To overcome this difficulty, we use a three level scheme, similar to the one used
in electromagnetically induced transparency (EIT) \cite{3level,EIT} where an optically thick substance is made
transparent and exhibits large dispersive response to the external field close to atomic resonance.

The particular EIT scheme here serves
several useful features required for left-handedness, such as being dispersive and  exhibiting resonance phenomena.
EIT materials do not suffer from linear absorption at resonance. They exhibit small transmission losses even
at high densities. As a consequence of the resonance, the EIT medium stores large amount of
energy over the cycles of interaction, leading to strong material response. In order to have
a negative electric and magnetic material response, we need both the
macroscopic polarization and magnetization of the material become simultaneously so strong  that
they would be immune even the sign changes of the applied fields. For the weak probe beam, EIT cannot
achieve this feat single-handedly. By considering a dense medium, with many particles within a cubic
resonance wavelength,
we let the local fields in the substance help to enhance the material responses. Indeed, we  see that,
circularly polarized probe electric field, under EIT conditions,
together with the help of the Lorentz-Lorenz local field contribution,
could maintain strong local currents that could give rise to large enough magnetization,
insensitive to sign changes of the probe magnetic field.
At the same time, the electric response also becomes negative.
The remainder of the section presents the mathematics behind these ideas as well as the conditions of their
applicability.

We require the three states to have the following non-zero matrix elements :
\begin{eqnarray}
\label{matrixcondition}
\langle 1 | e\vec{r} | 3 \rangle &\neq& 0 \\ \nonumber
\langle 2 | e\vec{r} | 3 \rangle &\neq&  0 \\ \nonumber
\langle 1 | \vec{\mu} | 2 \rangle &\neq& 0.
\end{eqnarray}
Here $\vec{\mu}$ is the magnetic dipole moment operator given by
\begin{equation}
\vec{\mu}= \frac{\mu_B}{\hbar} \left( g_L\vec{L} + g_S \vec{S} + g_I \vec{I} \right),
\end{equation}
where the first two terms are the magnetic moments due to the electronic orbital angular momentum $L$
and spin angular momentum $S$, while
the last term is the contribution of nuclear spin angular momentum $I$.
The coefficients are $g_L=1$ and $g_S=2$ (within a small $0.1\%$
correction found by quantum electrodynamical calculations). Nucleon magneton is about 1800 times
smaller than the Bohr magneton. Typical nuclear magnetic moments are about $1000$ times smaller
than their electronic counterparts and hence usually negligible. If the Hamiltonian is parity invariant,
we can choose all the states to be eigenstates of the parity operator ${\cal P}$.
To satisfy the requirement (\ref{matrixcondition}) one should have
\begin{equation}
\label{paritycondition}
\langle 1 | {\cal P} | 1 \rangle = \langle 2 | {\cal P} | 2 \rangle = -\langle 3 | {\cal P} | 3 \rangle.
\end{equation}

We assume that the states $|2\rangle$ and $|3\rangle$ are coupled with an intense coherent beam
while a weak probe beam will excite transitions between $|1\rangle$ and $|3\rangle$. We will investigate
the dielectric permittivity and magnetic permeability of a medium consisting of such atoms
as a response to the probe beam.

\begin{figure}
\centering{\vspace{0.5cm}}
\includegraphics[width=3.25in]{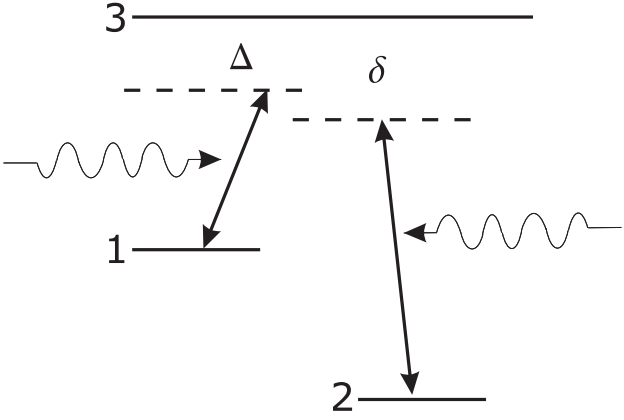}
\caption{Three-level atom interacting with the probe and the coupling fields in $\Lambda$ scheme
as described in the text.}
\label{fig1}
\end{figure}

Such a system of a three level atom interacting with those two optical fields in $\Lambda$ scheme
as depicted in Fig.\ref{fig1} is described by a Hamiltonian in the form
\begin{equation}
\mathrm{H} = \mathrm{H}_{0} + \mathrm{H}_{1},
\end{equation}
where
\begin{eqnarray}
{\mathrm H}_0 = \sum_{i=1}^{3}\hbar\omega_i R_{ii}
\end{eqnarray}
and
\begin{eqnarray}
{\mathrm H}_1 = -\frac{\hbar}{2}\sum_{i=1,2}\left(\Omega_i{\mathrm e}^{-{\mathrm i}\nu_it}R_{3i}+
{\mathrm c.c.}\right).
\end{eqnarray}
Here, $\hbar\omega_i$ are the energy levels of a free atom, and $R_{ij}=| i\rangle\langle j |$ are atomic
projection operators. The interaction Hamiltonian is written under the electric dipole approximation.
The Rabi frequencies associated with the optical transitions are defined by
\begin{eqnarray}
\Omega_i = \frac{\vec{d}_{3i}\cdot \vec{E}_i}{\hbar},
\end{eqnarray}
where the $\vec{E}_i$ stands for the complex amplitude of the positive frequency component electric
field of the probe laser. The electric dipole operator is expressed as
\begin{eqnarray}
\vec{d}_{3i} = e\langle 3 \mid \vec{r} \mid i \rangle.
\end{eqnarray}
Within the semiclassical theory of optical interactions, density matrix of the system evolves according
to the Liouville equation
\begin{eqnarray}
\frac{{\mathrm d}\rho}{{\mathrm d}t}=-\frac{{\mathrm i}}
{\hbar}[{\mathrm H},\rho]-\frac{1}{2}\{\Gamma,\rho\}.
\end{eqnarray}
Assuming a diagonal relaxation matrix $\langle i\mid\Gamma\mid j\rangle =\gamma_i\delta_{ij}$,
for our model Hamiltonian, density matrix equations (optical Bloch equations) become\cite{mori00,3level}
\begin{eqnarray}
\dot{\rho}_{33}&=&-\gamma_{3}\rho_{33}-\frac{{\mathrm i}}
{2}\sum_{i=1,2}\left(\Omega_i^\ast {\mathrm e}^{{\mathrm i}\nu_it}
\rho_{3i}-c.c.\right),\\
\dot{\rho}_{11}&=&-\gamma_{1}\rho_{11}-\frac{{\mathrm i}}{2}\left(\Omega_1
{\mathrm e}^{-{\mathrm i}\nu_1t}
\rho_{13}-c.c.\right),\\
\dot{\rho}_{22}&=&-\gamma_{2}\rho_{22}-\frac{{\mathrm i}}{2}\left(\Omega_2^\ast
{\mathrm e}^{{\mathrm i}\nu_2t}
\rho_{23}-c.c.\right),\\
\dot{\rho}_{31}&=&-({\mathrm i}\omega_{31}+\gamma_{31})\rho_{31}-\frac{{\mathrm i}}{2}\Omega_1
{\mathrm e}^{-{\mathrm i}\nu_1t}
(\rho_{33}-\rho_{11})\nonumber\\
&&+\frac{{\mathrm i}}{2}\Omega_2{\mathrm e}^{-{\mathrm i}\nu_2t}\rho_{21},\\
\dot{\rho}_{32}&=&-({\mathrm i}\omega_{32}+\gamma_{32})\rho_{32}-\frac{{\mathrm i}}{2}
\Omega_2{\mathrm e}^{-{\mathrm i}\nu_2t}
(\rho_{33}-\rho_{22})\nonumber\\
&&+\frac{{\mathrm i}}{2}\Omega_1{\mathrm e}^{-{\mathrm i}\nu_1t}\rho_{12},\\
\dot{\rho}_{21}&=&-({\mathrm i}\omega_{21}+\gamma_{21})\rho_{21}-\frac{{\mathrm i}}{2}
\Omega_1 {\mathrm e}^{-{\mathrm i}\nu_1t}
\rho_{23}\nonumber\\
&&+\frac{{\mathrm i}}{2} \Omega_2^\ast {\mathrm e}^{{\mathrm i}\nu_2t} \rho_{31},
\end{eqnarray}
where $c.c.$ implies the complex conjugate of the preceding term. It is useful to note here
that these equations
could be written more generally for a dense media in terms of the total local field. Within the linear
response theory, and assuming the material under consideration is linear, we will take into account
Lorentz-Lorenz correction after determining the dilute material response as usual\cite{bowd93}.

The relaxation rates of the off-diagonal elements of the density matrix are introduced as
$2\gamma_{ij}=\gamma_i+\gamma_j$.
Formal solution for $\rho_{21}(t)$ can be written as
\begin{eqnarray}
\rho_{21}(t)=\frac{{\mathrm i}}{2}\Omega_2^\ast {\mathrm e}^{{\mathrm i}\nu_2t}\int_0^\infty
\rho_{31}(t-t^{\prime}){\mathrm e}^{-{\mathrm i}[(\omega_{21}+\nu_2)+\gamma_{21}]t^{\prime}}
{\mathrm d}t^{\prime},\nonumber
\end{eqnarray}
which leads to
\begin{eqnarray}
\dot{\tilde{\rho}}_{31}&=&-({\mathrm i}\Delta+\gamma_{31})\tilde{\rho}_{31}-\frac{{\mathrm i}}{2}
\Omega_1 (\rho_{33}-\rho_{11})\nonumber\\
&&-\frac{\mid \Omega_2\mid^2}{4}\int_0^\infty
\tilde{\rho}_{31}(t-t^{\prime}){\mathrm e}^{-{\mathrm i}[(\Delta-\delta)+\gamma_{21}]t^{\prime}}
{\mathrm d}t^{\prime}.\nonumber
\end{eqnarray}
Here, we introduced a slow variable $\tilde{\rho}_{31}=\rho_{31}
{\mathrm e}^{{\mathrm i}\nu_1t}$, detuning of
the probe beam $\Delta = \omega_{31}-\nu_1$, and detuning of the driving beam
$\delta = \omega_{32}-\nu_2$.

The effect of weak probe field on the system can be treated perturbatively. Carrying out standard linear response
method, to the first
order in the probe field amplitude, we replace the inversion $(\rho_{33}-\rho_{11})$ by
its initial value which is taken to be $-1$, assuming only a small fraction of atoms are pumped out
of their initial states. The integral can be evaluated by assuming
$\tilde{\rho}_{31}$ doesn't change appreciably in time scale of $1/\gamma_{21}$.
We find
\begin{eqnarray}
\tilde{\rho}_{31} = \frac{{\mathrm i}}{2}\Omega_1
\frac{[{\mathrm i}(\Delta-\delta)+\gamma_{21}]}
{(i\Delta+\gamma_{31})[{\mathrm i}(\Delta-\delta)+\gamma_{21}]+\mid \Omega_2\mid^2/4}.
\end{eqnarray}
Positive frequency component of the complex induced electric dipole moment of the atom is given
by $p_i=d_{13}^{i}\rho_{31}$, which is related to the complex atomic polarizability tensor $\alpha$
as $p_i=\alpha_{ij} E_{1j}$. We adopt the summation convention, in which summation over a
repeated index is implied.

 For the macroscopic polarization we have to take into account local field
effects which lead to Clausius-Mossotti \cite{Jackson} relation between polarizability and the susceptibility $\chi_e$.
For small enough concentration $N$ of atoms $\chi_e=N\alpha\epsilon_0$ holds.
Using $P_i=\epsilon_0\chi_e^{ij} E_{1j}=d_{13}^{i}\rho_{31}$,  we identify the complex electric
susceptibility tensor for a gas of such three level atoms with concentration $N$ to be \cite{mori00,3level}
\begin{eqnarray}
\alpha^{ij} &=& \frac{{\mathrm i}}{2}\frac{d_{13}^i d_{31}^j}
{\gamma_{31}\hbar\epsilon_0}\frac{1}{D}\\
D &=& -\frac{\Delta}{\gamma_{31}}-{\mathrm i}
\left(1+\frac{\Omega_2^2}{4\gamma_{31}[{\mathrm i}(\Delta-\delta)+\gamma_{21}]}\right).
\nonumber\\
\chi_e&=&N\alpha\left(1-\frac{N}{3\epsilon_0}\alpha\right)^{-1}
\end{eqnarray}
Complex dielectric permittivity tensor can be similarly constructed via $\epsilon_{ij}
= \epsilon_0(\delta_{ij}+\chi_e^{ij})$.  We observe that this contributes
to the complex permeability tensor of the system.
It should be noted that for $\delta = 0$ and for small $N$, we recover the well-known results for an
electromagnetically induced transparent system. Now, using the equation
\begin{eqnarray}
\dot{\rho}_{21}=-({\mathrm i}\omega_{21}+\gamma_{21})\rho_{21}+\frac{{\mathrm i}}{2}
\Omega_2^\ast \tilde{\rho}_{31} {\mathrm e}^{{\mathrm i}(\nu_2-\nu_1)t},
\end{eqnarray}
we deduce the relation
\begin{eqnarray}
\tilde{\rho}_{21} = \frac{{\mathrm i}}{2}\frac{\Omega_2^\ast}
{{\mathrm i}(\Delta-\delta)+\gamma_{21}}\tilde{\rho}_{31},
\end{eqnarray}
for the new variable $\tilde{\rho}_{21} = \rho_{21}\exp{({\mathrm i}(\nu_1-\nu_2)t}$.

We can now calculate the induced magnetic dipole moment of the atom using
\begin{eqnarray}
\langle \vec{\mu} \rangle = Tr(\rho \vec{\mu}),
\end{eqnarray}
where $\vec{\mu}=\mu_B \vec{L}/\hbar$ is considered for the magnetic dipole operator by
assuming the contribution from nuclear spin are negligible. The electronic spin part is
for simplicity.

As the lower levels are of opposite parity with the upper level, the only non-vanishing contribution
may arise if the lower levels are of the same parity. In this case we get
$\langle \vec{\mu} \rangle=\rho_{21}\vec{\mu}_{12}+c.c$, which gives
\begin{eqnarray}
\langle \vec{\mu} \rangle = -\frac{\Omega_1\Omega_2^\ast\vec{\mu}_{12}
\exp{({\mathrm i}(\nu_1-\nu_2)t)}}
{4({\mathrm i}\Delta+\gamma_{31})[{\mathrm i}(\Delta-\delta)+\gamma_{21}]+\mid \Omega_2\mid^2}
+c.c.
\end{eqnarray}

In order to describe the atomic response to the magnetic field component of the probe
field, we let the induced magnetic dipole of the atom oscillate in phase with the probe
beam. This is achieved when $\nu_1-\nu_2 = \pm \nu_1$. Setting aside the static field solution
we consider the case of $\nu_2=2\nu_1$. The other possibility $\nu_2=0$ would be the case of
a static electric field as the coupling field. This should be separately discussed as it is
necessary to examine Stark shifts of the levels and modify the present theory accordingly.
The driving field is taken to be resonant with the $\omega_{32}$
when the probe is resonant with the $\omega_{31}$ so that $\delta = 2\Delta$ which puts a constraint
on the three level system as $\omega_{32}=2\omega_{31}$.

This constraint is, however, a major obstacle in realizing the predicted effects here at
a realistic experimental setting as it is not straightforward to find a system with two states, which have a matrix
element of $\vec{\mu}$ between them and at the same time have energy difference in the optical range.
This is mainly due to the fact that $\mu$ is an angular operator and the two states involved should have the
same radial wavefunctions to give a non zero matrix element. One can imagine, some external magnetic field
adjusting the separations to give the necessary energy conditions. However, for an atomic system to get splittings
in the optical regime, the external field would be impractically large. One can try to investigate systems
in which $\vec{\mu}$ is not an angular operator,
such as molecular gases, and try to find  optically separated states which have a magnetic
dipole matrix element between them.

As far as atomic gases are concerned, the best option seems to be to take two states which have the same $L$ value
but which are split due to L-S coupling to be the states $|1 \rangle$ and $| 2 \rangle$ and try to get a
third level of opposite parity to fulfill the energy condition.
Another direction to proceed would be to consider all our discussion for an atomic
system under high electric field. In that case, it will not be too hard to get to fields which give shifts on the
order of optical frequencies, however one must carefully do the preceding analysis again taking into account
the effect of static electric field on all three states.

We assume this condition is fulfilled
with our hypothetical model atom and proceed by
writing the product $\Omega_1\vec{\mu}_{12}$ explicitly, so that we can examine
the directional character of the
magnetic response of the atom to the probe field. Electric dipole of
the probe transition and the magnetic dipole of the lower levels are combined through a tensor product
relation such that
\begin{eqnarray}
\Omega_1\left(\vec{\mu}_{12}\right)_i&=&\frac{E_1}{\hbar}\sum_j\nu^{ij}\epsilon_{1j},
\end{eqnarray}
where we introduce
\begin{eqnarray}
\nu^{ij}=\langle 1|\mu_i|2\rangle\langle 3|d_{31}^j|1\rangle.
\end{eqnarray}
The tensor $\nu$ demonstrates the combined effect of electric and magnetic field components
of the optical field on the directional character of the magnetic response of the medium.

To calculate the induced magnetic dipole moment matrix elements, it is convenient to consider
angular momentum basis in which we can also calculate the elements of electric dipole
moment using the Wigner-Eckart theorem. Let us identify the states as
\begin{eqnarray}
|1 \rangle &\doteq& | n , l , m\rangle, \nonumber\\
|2 \rangle &\doteq& | n, l , m-1 \rangle, \\
|3 \rangle &\doteq& | n', l+1 , m-1 \rangle. \nonumber
\end{eqnarray}
Using $L_x=(L_++L_-)/2$ and $L_y=(L_+-L_-)/2{\mathrm i}$ matrix elements of the angular momentum are
readily obtained in this basis as
\begin{eqnarray}
\langle 1|L_x|2\rangle &=&\frac{\hbar}{2}\sqrt{(l+m)(l-m+1)},  \nonumber \\
\langle 1|L_y|2\rangle &=&\frac{\hbar}{2{\mathrm i}}\sqrt{(l+m)(l-m+1)}, \nonumber\\
\langle 2|L_z|1\rangle &=& 0.
\label{eq:mdipole}
\end{eqnarray}

The matrix elements of the electric dipole operator can be conveniently calculated by expressing
it as a spherical tensor operator of rank $1$ so that its components become
\begin{eqnarray}
ez &=& T^{(1)}_0;\quad ex = \frac{1}{\sqrt{2}}\left(T_{-1}^{(1)}-T_1^{(1)}\right); \nonumber\\
ey &=& \frac{{\mathrm i}}{\sqrt{2}}\left(T_{-1}^{(1)}+T_1^{(1)}\right).
\end{eqnarray}
We use the Wigner-Eckart theorem in the form
\begin{eqnarray}
\langle nl_3m_3|T^{(l_2)}_{m_2}|n^{\prime}l_1m_1\rangle=C^{l_1l_2l_3}_{m_1m_2m_3}
\frac{\langle nl_3|| T^{(l_2)}||n^{\prime}l_1\rangle}{\sqrt{2l_1+1}}.\nonumber
\end{eqnarray}
Here, the first factor is the Clebsh-Gordan coefficient where we choose a notation resembling
its symmetric form in terms of Wigner-3j coefficients. The second factor is the
reduced matrix element which is independent of the orientation of the magnetic dipole characterized
by the angular momentum projection quantum number $m$. In our case it is given by
\begin{equation}
\langle n^{\prime},l+1| |e\vec{r}| |n,l \rangle = \int_0^\infty {\mathrm d}r
er^3 R^*_{n^{\prime},l+1}(r) R_{nl}(r),
\end{equation}
which is always non-vanishing, with $R_{nl}(r)$ being the radial wavefunction.
By the $m$-selection rule ($m_1+m_2=m_3$ is
required for non-vanishing matrix elements), we see that matrix
elements of the $T^{(1)}_{1,0}$ vanish. The sole nonvanishing matrix element of $T^{(1)}_{-1}$ determines
the matrix elements of $x$ and $y$ components of the position operator which are found to be
\begin{eqnarray}
d_{31}^x&=&
-\sqrt{\frac{(l-m+1)(l-m+2)}
{2(2l+2)(2l+3)^3}}
\langle n^{\prime},l+1||e\vec{r}||nl\rangle,\nonumber\\
d_{31}^y&=&id_{31}^x,\quad d_{31}^z = 0.
\label{eq:edipole}
\end{eqnarray}
Combining Eq.\ref{eq:mdipole} and Eq.\ref{eq:edipole}, we finally get
\begin{eqnarray}
\nu &=& \frac{\mu_B}{4} \langle n^{\prime},l+1| |e\vec{r}| |nl \rangle
(l-m+1)\\ \nonumber
&\times& \sqrt{\frac{(l+m)(l-m+2)}{(l+1)(2l+3)^3}} \left[ \begin{array}{ccc}
-1 &  -i & 0\\
 i &  -1 & 0\\
 0 &   0 & 0 \end{array} \right]
\end{eqnarray}
as the matrix which determines the orientation of the induced dipole moment.
It should be noted that the matrix $\nu$ gives zero response to
positively  polarized EM waves in accordance with the dipole selection rules.
For our particular set of levels, we need
negatively  polarized EM waves as they provide the photons with correct helicity
to satisfy the angular momentum conservation in the
probe photon emission and absorption processes between the states
1 and 3. For negatively polarized waves, $\nu$ just
reduces to a scalar.

It is worth noticing that the structure of the tensor $\nu$ resembles that of gyrotropic substances
with both $\epsilon$ and $\mu$ are tensors such as pure ferromagnetic metals and semiconductors. These
were argued to be most likely candidates to demonstrate left-handedness in the original paper by
Veselago\cite{Veselago}.

Further calculations require setting the polarizations of the coupling and the probe beams. To cause
transitions between $|2\rangle$ and $|3\rangle$;
the coupling beam polarization $\hat{\epsilon}_d$ has to have a component along the quantization direction
$\hat{z}$. So let us take the coupling beam to propagate in the $x-y$ plane, and be linearly polarized along
$\hat{z}$. To cause transitions between states $|1\rangle$ and $|3\rangle$,
the probe beam must have a polarization vector lying in the $x-y$ plane. Let us take it to be propagating along
the $\hat{z}$ axis with polarization lying in the x-y plane.

Then, our general expression for the induced magnetic moment leads to
\begin{equation}
\vec{\mu} = \gamma(\omega) E (\hat{x} - i \hat{y}),
\end{equation}
where $\omega$ now denotes the frequency of the probe beam and
\begin{eqnarray}
\gamma(\omega) &=& \frac{\mu_B}{2\hbar}\Omega_2^\ast \langle n^{\prime},l+1| |e\vec{r}| |nl \rangle
(l-m+1)\nonumber\\
&&\times\sqrt{\frac{(l+m)(l-m+2)}{(l+1)(2l+3)^3}}\frac{1}{Z}\\
Z&=&4({\mathrm i}\Delta+\gamma_{31})[{\mathrm i}(\Delta-\delta)+\gamma_{21}]+\mid \Omega_2
\mid^2\nonumber
\end{eqnarray}
With this definition of $\gamma(\omega)$ we can extend our result to macroscopic electromagnetics
of a gas with concentration N. In the spirit of Clausius--Mossoti equation \cite{Jackson}, we define the
magnetization per unit volume as
\begin{eqnarray}
\label{Mequation}
\vec{M} &=& N \gamma(\omega) ( E + \frac{P}{3\epsilon_0} ) (\hat{x} - i \hat{y})\\ \nonumber
&=& N  \gamma(\omega) (1+\frac{\chi_e}{3}) E (\hat{x} - i \hat{y}).
\end{eqnarray}
Now we  recall the Fourier transform of the curl equation for electric field
in Maxwell's equations. For a negatively polarized wave
\begin{eqnarray}
\vec{B}&=& \frac{1}{\omega}\vec{k} \times E (\hat{x} - i \hat{y})\nonumber\\
&&=\frac{i}{c}E(\hat{x} - i \hat{y}).
\label{Bequation}
\end{eqnarray}
Combining equations (\ref{Mequation}),(\ref{Bequation}), we have
\begin{equation}
\vec{M} = -iN \gamma(\omega)c (1+\frac{\chi_e}{3})\vec{B}.
\end{equation}
Finally by using the definitions $ \vec{B} = \mu_0(\vec{H} +\vec{M}) $ and $\vec{B} = \mu \vec{H}$, we get
\begin{equation}
\mu_r(\omega) = \frac{1}{1 + i\mu_0\gamma(\omega)c (1+\frac{\chi_e(\omega)}{3})}
\end{equation}
with $\mu_r=\mu/\mu_0$ is the relative (effective)
permeability. We shall see that combined effect of electric and magnetic
field components of the optical fields, as well as local field effects lead to novel light propagation regimes
in particular on the EIT resonance frequency.
\section{Results and Discussions}

\begin{figure}
\centering{\vspace{0.5cm}}
\includegraphics[width=3.25in]{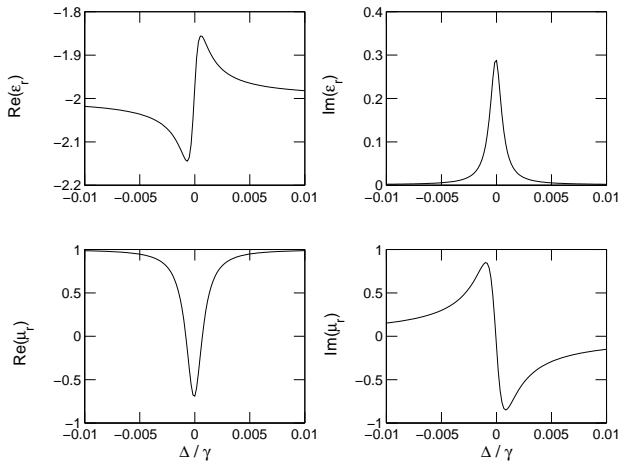}
\caption{Frequency dependence of the relative (effective)
dielectric permittivity $\epsilon_{\mathrm r}$ and the relative
magnetic permeability $\mu_{\mathrm r}$ of the dense gas of three level atoms with $N=10^{24}{\mathrm m}^{-3}$,
$\lambda \sim 589 \mathrm {nm}$, $\gamma\sim 10.06 \mathrm {MHz}$, $\gamma_{ge}=0.5\gamma$,
$\gamma_{gr}/2\pi \sim 10^3 \mathrm {Hz}$, and $\Omega_2=0.56\gamma$.}
\label{fig2}
\end{figure}

\begin{figure}
\centering{\vspace{0.5cm}}
\includegraphics[width=3.25in]{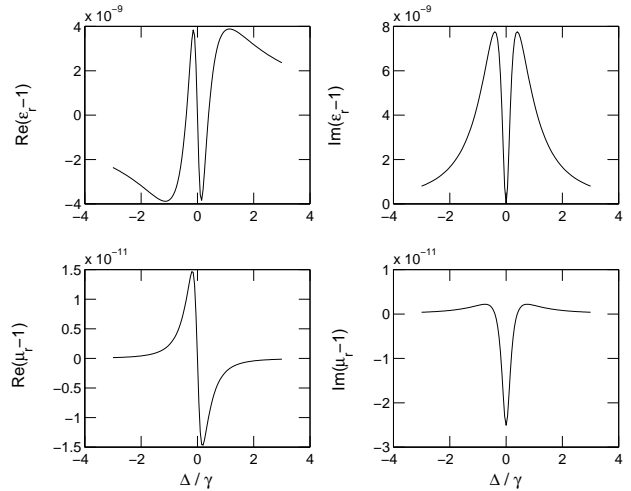}
\caption{Same with Fig.\ref{fig2} but for the case of dilute gas with $N=10^{12} {\mathrm m}^{-3}$.
Here, the electric susceptibility and the magnetic susceptibility are plotted as the relative
permittivity and the permeability do not change appreciably from unity.}
\label{fig3}
\end{figure}

We consider a gas of  $^{23}$Na atoms with
$N=10^{24}{\mathrm m}^{-3}$ to examine the case of dense media where the Lorentz-Lorenz local
field corrections play significant role and $N=10^{12} {\mathrm m}^{-3}$,
for the case of a dilute gas where the local field effects are weak. Our results are presented in
Fig.\ref{fig2} for the dense media and in Fig.\ref{fig3} for the dilute gas.

We see that both the relative dielectric permittivity $\epsilon_{\mathrm r}=\epsilon/\epsilon_0$ and
the relative magnetic permeability $\mu_{\mathrm r}=\mu/\mu_0$ can become negative over a band of
frequency $\sim 0.001\gamma$. This allows the propagation of light through otherwise opaque
medium at high densities where the electromagnetically induced transparency would not work.
At resonance we find $\mu_r(0)=-0.69-{\mathrm i}0.11$ and $\epsilon_r(0)=-1.86+{\mathrm i}0.12$.
It is natural to have transmission losses in our model, similar to other left-handed structures,
as they are unavoidable due to the Kramers-Kronig relations ensuring the causality in the system.
On the other hand, theoretically it is not a trivial task to estimate the amount
of losses\cite{garc02,mark02,para03}
and to rigorously prove the causality in left-handed materials\cite{ziol03,kosc03}.
We can give a simple and rough estimate by simply taking into account the imaginary
part of the refractive index which gives that after several microns the optical field will be
damped by $\sim 33\%$ due to linear absorption. At such length scales, our atomic system with
the given densities may be found in Bose-Einstein condensed state due to the interatomic
interactions. Multiple scattering of photons
as well as higher order many body correlations may contribute in addition to the local field correction.
Such effects are argued to be about the same order with the local field correction
\cite{krut99,wall97,mori95,ruos97_1,ruos97_2,flei99,krut00}. It is an intriguing
possibility that the present result of induced left-handedness could improve and benefit
from contributions arising from the quantum correlations in a dense Bose-Einstein condensate or in a
dense degenerate Fermi gas. In this paper, we will be content with limiting ourselves
to classical gaseous media and hope to discuss the case of quantum gases elsewhere in detail.

In the dilute gas limit, we recover the usual behavior of the electric susceptibility under
electromagnetically induced transparency conditions. The transparency region is in the valley between the
twin peaks in the imaginary part of the electric susceptibility where the peaks correspond to the
two dressed absorption lines, the Autler-Townes doublet\cite{autl55}. The magnetic susceptibility exhibits steep
variation over a narrow band of frequencies in the vicinity of the resonance while in magnitude
the relative permeability remains close to unity for all $\Delta$. When $N\sim 10^{20}$, similar
results to those shown in Fig.\ref{fig3} are found where now $\mu_r$ varies between $1.002$ and
$0.9985$ over $\Delta \in (-\gamma,\gamma)$.

In our numerical calculations, we estimate the dipole matrix element from the spontaneous
emission rate $\gamma\sim 10.06 \mathrm {MHz}$ using the relation
$d_{31} = \sqrt{3\gamma\hbar\epsilon_0\lambda^3/8\pi^2}$. Here $\lambda$ is the wavelength of resonant
probe transition which is $\lambda \sim 589 \mathrm {nm}$.
Typical values for $\gamma_{ge}=0.5\gamma$ and $\gamma_{gr}/2\pi \sim 10^3 \mathrm {Hz}$ are used.
Rabi frequency associated with the driving field is chosen to be $\Omega_2=0.56\gamma$.

\label{sec:results}
\section{Conclusion}
In summary, we  suggested a method for optical modification
of magnetic permeability using a three level scheme and derived the necessary conditions for its applicability.
We  found that it is in principle possible to electromagnetically
induce left-handedness to a spatially homogeneous  media. The
major challenge we face is to have two levels separated at optical frequencies
while having a non-vanishing magnetic dipole matrix element. Such level splittings require large
external magnetic fields or should be engineered by other means such as external electric fields or spin-orbital
couplings. One may also consider molecular gases, or try to utilize excitonic energy levels in
solid state heterostructures to engineer three level system fulfilling the energy condition.
The predicted effect is fundamentally based upon the Lorentz-Lorenz local field contribution in an
electromagnetically induced transparency medium of three level atoms with a non-vanishing dipole moment between
lower levels. In dense medium limit, in which the medium becomes opaque normally with a negative dielectric
constant, the presence of magnetic dipole gives rise to a negative magnetic permeability so that the
probe beam would still propagate within the otherwise optically thick dense medium for several microns before
it is finally absorbed.

It should be emphasized that the presented method is applicable to spatially
homogeneous media and does not need any spatial periodicity which is unavoidable in metamaterials.
In the dilute medium limit, the value of permeability do not change from unity appreciably, however, in this
case we observed that it demonstrates steep changes over a small band of frequency. Such large gradient of
permeability may affect the character of light propagation such as its group velocity and may serve an additional
method to slow down or speed up the light. \label{sec:conclusion}


\acknowledgments

O.E.M. acknowledges useful discussions with A. Sennaroglu.
M.O.O. thanks W. Ketterle for a preliminary discussion of the idea and encouragement.


\begin{thebibliography}{}

\bibitem{slow}
L.V. Hau, S.E. Harris, Z. Dutton, and C.H. Behroozi, Nature {\bf 397}, 594 (1999).

\bibitem{stop}
D.F. Phillips, A. Fleischhauer, A. Mair, R.L. Walsworth, and M.D. Lukin,
Phys. Rev. Lett. {\bf 86}, 783 (2001).

\bibitem{physicstoday}
J.B. Pendry and D.R. White, Physics Today, {\bf 57}(6), 37 (2004)
and references therein.

\bibitem{focusing}
J.B. Pendry, Phys. Rev. Lett. {\bf 85}, 3966 (2000).

\bibitem{quantumcomputation}
A. Mair, J. Hager, D.F. Phillips, R.L. Walsworth, and M.D. Lukin,
Physical Review A 65, 031802 (2002).

\bibitem{Jackson}
J.D. Jackson, {\it Classical Electrodynamics}, 2nd ed., (Wiley, New York, 1975).

\bibitem{3level}
M.O. Scully and M.S. Zubairy, {\it Quantum Optics},
(Cambridge University Press, Cambridge, 1997).

\bibitem{Landau}
L.D. Landau, E.M. Lifshitz and L.P. Pitaevskii,
{\it Electrodynamics of Continuous Media},
(Butterworth Heniemann, Oxford, 1998).

\bibitem{Veselago}
V.G. Veselago, Usp. Fiz. Nauk. {\bf 92}, 517 (1964);
Translation in Sov. Phys. Usp. {\bf 10}, 509 (1968).

\bibitem{leftexperiment}
D.R. Smith, W.J. Padilla, D.C. Vier, S.C. Nemat-Nasser, and S. Schultz,
Phys. Rev. Lett. {\bf 84}, 4184 (2000).

\bibitem{lefttheory}
J.B. Pendry, A.J. Holden, W.J. Stewart and I. Youngs,
Phys. Rev. Lett. {\bf 76}, 4773 (1996).

\bibitem{ertugrul}
E. Cubukcu, K. Aydin, E. Ozbay, S. Foteinopoulou, and C.M. Soukoulis,
Phys. Rev. Lett. {\bf 91}, 207401 (2003).

\bibitem{EIT}
S. E. Harris, J. E. Field, and A. Imamo\u{g}lu, Phys. Rev. Lett. {\bf 64}, 1107 (1990);
K.J. Boller, A. Imamo\u{g}lu, and S.E. Harris, Phys. Rev. Lett. {\bf 66}, 2593 (1991);
M. O. Scully, Phys. Rep. {\bf 219}, 191 (1992);
S. E. Harris, Physics Today {\bf 50}, 36 (1997), and references therein.

\bibitem{mori00} G. Morigi and G.S. Agarwal, Phys. Rev. A {\bf 62}, 013801 (2000).

\bibitem{bowd93} C. M. Bowden and J. P. Dowling, Phys. Rev. A {\bf 47}, 1247 (1993).

\bibitem{garc02} N. Garcia and M. Nieto-Vesperinas, Optics Letters {\bf 27}, 885 (2002).

\bibitem{mark02} P. Marko\v{s}, I. Rousochatzakis, and C. M. Soukoulis, Phys. Rev. E {\bf 66}, 045601 (2002).

\bibitem{para03} C. G. Parazzoli, R. B. Greegor, K. Li, B. E. C. Koltenbah, and M. Tanielian,
Phys. Rev. Lett. {\bf 90}, 107401 (2003).

\bibitem{ziol03} R. W. Ziolkowski and A. D. Kipple, Phys. Rev. E {\bf 68}, 026615 (2003).

\bibitem{kosc03} T. Koschny, P. Marko\v{s}, D. R. Smith, and C. M. Soukoulis,
Phys. Rev. E {\bf 68}, 065602 (2003).

\bibitem{krut99} K. V. Krutitsky, F. Burgbacher, and J. Audretsch, Phys. Rev.
A {\bf 59}, 1517 (1999).

\bibitem{wall97} H. Wallis, Phys. Rev. A {\bf 56}, 2060 (1997).

\bibitem{mori95} O. Morice, Y. Castin, and J. Dalibard, Phys. Rev. A {\bf 51}, 3896
(1995).

\bibitem{ruos97_1} J. Ruostekoski and J. Javanainen, Phys. Rev. A {\bf 55}, 513 (1997).

\bibitem{ruos97_2} J. Ruostekoski and J. Javanainen, Phys. Rev. A 56, 2056
(1997).

\bibitem{flei99} M. Fleischhauer and S. F. Yelin, Phys. Rev. A {\bf 59}, 2427 (1999).

\bibitem{krut00} K. V. Krutitsky, F. Burgbacher, and J. Audretsch, Laser Phys. {\bf 10}, 15 (2000).

\bibitem{autl55} S. H. Autler, C. H. Townes, Phys. Rev. {\bf 100}, 703 (1955).

\end{thebibliography}
\end{document}